\documentclass[aps,prc,twocolumn,tightenlines,showpacs,
showkeys,superscriptaddress,nofootinbib]{revtex4}
\usepackage{graphicx}
\usepackage[figuresright]{rotating}

\begin{document}


\title{New Outlook on the Possible Existence of Superheavy Elements in
Nature\footnote{Paper presented at VII Int. School-Seminar on
Heavy Ion Physics, May 27 - June 1, 2002, Dubna Russia. } }


\author{A. Marinov}
\email{marinov@vms.huji.ac.il}
\author{S. Gelberg}
\affiliation{The Racah Institute of Physics, The Hebrew
University, Jerusalem 91904, Israel}
\author{D. Kolb}
\affiliation{Department of Physics, University GH Kassel, 34109
Kassel, Germany}
\author{R. Brandt}
\affiliation{Kernchemie, Philipps University, 35041 Marburg,
Germany}
\author{A. Pape}
\affiliation{IReS-UMR7500, IN2P3-CNRS/ULP, BP28, F-67037
Strasbourg cedex 2, France}


\date{July 23, 2002}

\begin{abstract}
A consistent interpretation is given to some previously
unexplained phenomena seen in nature in terms of the recently
discovered long-lived high spin super- and hyper-deformed isomeric
states. The Po halos seen in mica are interpreted as due to the
existence of such isomeric states in corresponding Po or nearby
nuclei which eventually decay by $\gamma$- or $\beta$-decay to the
ground states of $^{210}$Po,$^{214}$Po and $^{218}$Po nuclei. The
low-energy 4.5 MeV $\alpha$-particle group observed in several
minerals is interpreted as due to a very enhanced
 $\alpha$ transition from the third minimum of the potential-energy
 surface in  a
superheavy nucleus with atomic number Z=108 (Hs) and atomic mass
number around 271 to the corresponding minimum in the daughter.
\end{abstract}

\maketitle

\section{Introduction}
The theoretical predictions in the  nineteen sixties
\cite{str64,mye66,sob66,str67,won67,muz68,nil68,gru69} of the
possible existence of very long-lived superheavy elements around
Z=114 and N=184 stirred a lot of excitement among the nuclear
scientific community and have initiated the search for the
possible existence of superheavy elements in nature.
 In the present paper we
concentrate on two independent well established  experimental
results which are impossible to understand  under the present
common knowledge of nuclear physics. These puzzling data are first
the observation, in mica minerals, of certain halos which have
been attributed to the $\alpha$-decay of the short-lived
$^{210}$Po,$^{214}$Po and $^{218}$Po nuclei
\cite{hen39,gen68,gen92}, and secondly the observation in several
minerals of a low energy $\alpha$-particle group with an energy of
about 4.5 MeV \cite{cher63,che64,cher68,mei70}.

Halos in mica, which consist of tiny concentric  rings, are known
for a long time \cite{jol07,mug07}. For most of them the measured
radii of the rings fit with the known ranges of the various
$\alpha$-particle groups from $^{238}$U or $^{232}$Th decay
chains. Therefore they have been correctly interpreted, back in
1907 \cite{jol07,mug07}, as due to the existence of very small
grains of $^{238}$U or $^{232}$Th in the centers of the
corresponding halos, which have been decaying, through their
characteristic decay chains, since the time of crystallization of
the crust of the earth. However, other types of halos have been
discovered back in 1939 \cite{hen39} and have been thoroughly
studied
 by Gentry \cite{gen68}. These are the $^{210}$Po,
$^{214}$Po and $^{218}$Po halos which consist of respective one,
two and three concentric rings, with radii equal to the ranges of
the $\alpha$-particle groups from the decay chains of the
corresponding Po isotopes\footnote{Colored pictures of various
halos are given in Ref. \cite{gen92}.}. These Po isotopes belong
to the $^{238}$U decay chain. However, their half-lives, as well
as the half-lives of their $\beta$-decay parents, are short, and
since rings belonging to their long-lived precursors are absent,
their appearance in nature is puzzling \cite{fea78}.

Another puzzling phenomenon is the  low-energy $\alpha$-particle
group, around 4.5 MeV, which has been seen in molybdenite
\cite{cher63}, in thorite \cite{che64}, in magnetite \cite{cher68}
and in OsIr \cite{mei70}. The cleanest spectrum, where this group
appears without observed residues from U isotopes decays, was
obtained by Cherdyntsev et al. \cite{cher68}. Based on chemical
behavior (having volatile oxides) of the $\alpha$-emitter, it has
been suggested that it might be due to a decay of an isotope of
Eka-Os, the superheavy element with Z=108 (Hs).\footnote{Actually
Cherdyntsev suggested naming element 108 sergenium, based on part
of the great silk path in Kazakhstan (name Serika) where the
studied  mineral molibdenite was found (private communication from
Yu. Lobanov.)} Since it was usually found together with $^{247}$Cm
and $^{239}$Pu, it has been suggested \cite{mei70} that this
low-energy $\alpha$-particle group is due to an isotope of element
108  which is a precursor of $^{247}$Cm and its descendant
$^{239}$Pu. The half-life of this activity has been estimated to
be around $(2.5\pm0.5)\times10^8$ y \cite{cher63}.

With the current common knowledge of nuclear physics it seemed
impossible to understand these data. The predicted energies of
ground state to ground state $\alpha$ transitions for $\beta$
stable Hs nuclei with atomic masses of 274 to 286 are between 9.5
and 6.7 MeV \cite{mnk,kuty,lmz}, and the predicted  half-lives for
these energies are between $3\times10^{-2}$ s and $3\times10^2$ y
\cite{vs66,roy00}, as compared to half-life of about
$5\times10^{16}$ y for 4.5 MeV. The question is why the nucleus
decays with such a low energy $\alpha$-particle when a much higher
energy, with a penetrability factor of at least 14 orders of
magnitude higher, is available.

A second question is how the nucleus can decay with about eight
orders of magnitude shorter lifetime than what is predicted
\cite{vs66,roy00} from energy versus lifetime relationships for a
normal 4.5 MeV  $\alpha$-transition (experimentally estimated
$2.5\times10^{8}$ as compared to predicted $5\times10^{16}$ y).
 (A lifetime in the region of $10^{16}$ y is
certainly impossible, since it implies the existence of about 100
mg of material in the studied samples).

In the following paragraphs similar effects  seen in the study of
various actinide fractions \cite{mar01b} produced via secondary
reactions \cite{mar71}, and also in the study of the
$^{16}$O+$^{197}$Au \cite{mar96a,mar96b} and the
$^{28}$Si+$^{181}$Ta \cite{mar01a} heavy ion reactions, are
summarized. Based on the results of all these experiments, a
consistent interpretation for the puzzling phenomena seen in
nature is suggested. (See also Ref. \cite{mar02}).

\begin{widetext}

\begin{figure}[h]
\includegraphics[width=1.0\textwidth,angle={-1.0}]{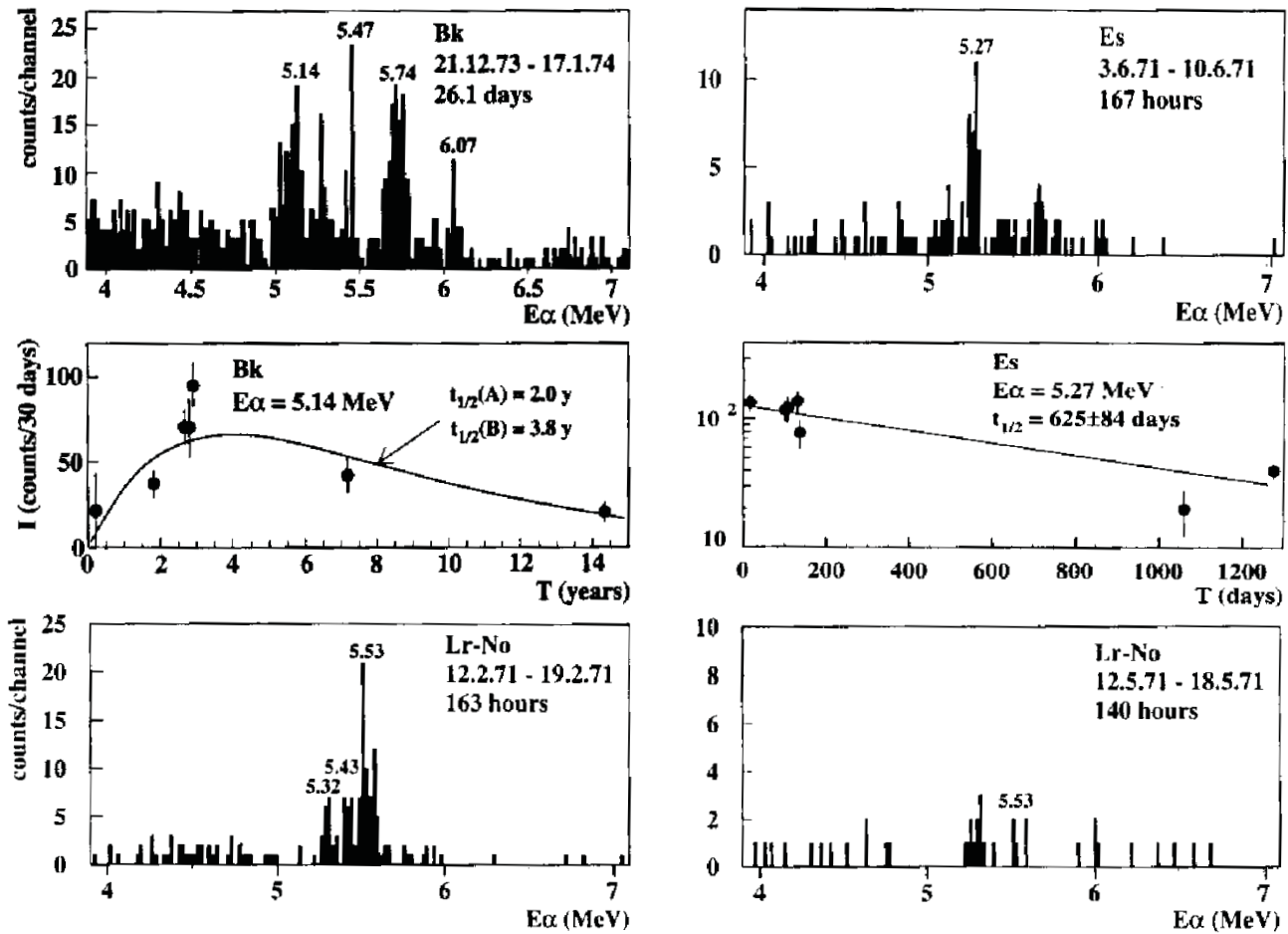}
\vspace*{-0.9cm} \caption{{\bf Left, top:} $\alpha$-particle
spectrum obtained with the Bk source. {\bf Right, top:}
$\alpha$-particle spectrum obtained with the Es source. {\bf Left,
center:} Decay curve obtained with the 5.14 MeV group seen with
the Bk source (left, top above).(See Comment (a) in Table~II
regarding the growing half-life of 2.0 y \cite{mar01b}). {\bf
Right, center:} Decay curve obtained with the 5.27 MeV group seen
with the Es source (right, top above). {\bf Left, bottom:}
$\alpha$-particle spectrum obtained with the Lr-No source. {\bf
Right, bottom:} The same as the previous one but taken about 3
months later. From a comparison of the two spectra a half-life of
26$\pm$7 d was deduced for the 5.53 MeV group \cite{mar01b}.}
\end{figure}

\end{widetext}

\section{Unidentified $\alpha$-particle groups in actinides}

 In a study of actinide
fractions from a W target which had been irradiated with 24-GeV
protons, long-lived isomeric states were found in the
neutron-deficient $^{236}$Am and $^{236}$Bk nuclei with respective
half-lives of 0.6~y and $\geq$30 d \cite{mar87}. Their character
however was not clear, being far from closed shell nuclei, where
high spin isomers are known, and living much longer than the known
fission isomers. In addition, several unidentified
$\alpha$-particle groups were found in some actinide sources.
Thus, 5.14 MeV (t$_{1/2}$ = 3.8$\pm$1 y),
  5.27 MeV  (t$_{1/2}$
= 625$\pm$84 d) and  5.53 MeV (t$_{1/2}$ = 26$\pm$7 d) groups were
respectively found in the Bk, Es and Lr-No sources
\cite{mar87,mar01b} (See Fig. 1 and Table I). Similar to the
situation with the 4.5 MeV group seen in nature, also here in the
case of the latter unidentified groups, one can not understand
their relatively low energies  (e.g., 5.53 MeV in Lr-No as
compared to typical g.s. to g.s. transitions of around 8 MeV
(Column 6 of Table I), which have about 11 orders of magnitude
larger penetrability factors (the ratio of Column 4 to Column 7 in
Table I)), and their very enhanced character, having 10$^{5}$ -
10$^{7}$ shorter half-lives than predicted from
 the systematics  of energy versus half-life
 relationship for normal $\alpha$-decays \cite{vs66,roy00}
  (See Column 5 in Table I).
 The deduced evaporation-residue cross sections
\cite{mar01b}, in the mb region, are also several orders of
magnitude larger than expected.

\begin{table}[h]
\vspace*{-0.5cm} \caption[]{The energies and half-lives of several
unidentified $\alpha$-particle groups seen in various actinide
sources as compared to theoretical predictions.}
\begin{minipage}{0.5\textwidth} 
\renewcommand{\footnoterule}{\kern -3pt} 
\begin{tabular}{lllllll}
\hline\\[-10pt]         
Source &     E$_{\alpha}^{exp}$   &   t$_{1/2}^{exp}$   &
t$_{1/2}^{cal}$\cite{roy00}   &  Enhance-   &
E$_{\alpha}$\footnote{Typical values from Ref. \cite{aud93}.} &
t$_{1/2}^{cal}$\footnote{Calculated for the lower energy of Column
6 according to formulas given in Ref. \cite{roy00}.} \\
   &     &    &     &  ment  &  g.s.$\rightarrow$g.s.  &    \\
   &  (MeV)  &  (y)  &  (y) &  factor\footnote{The ratio of Column 4 to Column 3.}
    &  (MeV)  &  (s)\\
\hline\\[-10pt]
  Bk  &  5.14   &  3.8  &  1.7$\times10^{5}$\footnote{Calculated
  for $^{238}$Am. See below.}  &  4.5$\times10^{4}$  &  6 - 7  &
 2.2$\times10^{9}$\\
  Es  &  5.27   &  1.7  &
  2.7$\times10^{6}$\footnote{Calculated for $^{247}$Es. See below.}   &
   1.6$\times10^{6}$  &
  7 - 8  &  1.9$\times10^{6}$\\
  No-Lr  &  5.53   &  0.07  &
  1.1$\times10^{6}$\footnote{Calculated for $^{252}$No. See below.}
   &  1.5$\times$10$^{7}$
  &  8 - 9  &  2.4$\times10^{2}$\\

\hline
\end{tabular}
\end{minipage}
\renewcommand{\footnoterule}{\kern-3pt \hrule width .4\columnwidth
\kern 2.6pt}            
\end{table}
\vspace*{-0.5cm}

\section{Study of the $^{16}$O+$^{197}$A\lowercase{u} reaction and long-lived
high spin superdeformed isomeric states}

Possible explanation for the above puzzling data comes from the
study of the $^{16}$O+$^{197}$Au reaction at E$_{Lab}$=80 MeV
which is around the Coulomb barrier \cite{mar96a,mar96b}, and of
the $^{28}$Si+$^{181}$Ta reaction at E$_{Lab}$=125 MeV
\cite{mar01a}, about 10\% below the Coulomb barrier. In the first
reaction a 5.2 MeV $\alpha$-particle group with a half-life of
about 90 m has been found in $^{210}$Fr. This group has the same
unusual properties as the abnormal $\alpha$-particle groups found
in the actinides and produced via secondary reactions,   and  of
the 4.5 MeV found in nature. 5.2 MeV is a low energy as compared
to 6.5 MeV, the g.s. to g.s. transition from $^{210}$Fr, and 90 m
half-life  is about 4$\times$10$^{5}$ enhanced as compared to the
prediction \cite{vs66,roy00} for normal $\alpha$-particles of this
energy from $^{210}$Fr. However, the 5.2 MeV group has been found
in coincidence with $\gamma$-rays which fit the energies of a
superdefermed band transitions. Therefore, the $\alpha$-decay is
through a barrier of a superdeformed nucleus, and the large
enhancement can quantitatively be understood \cite{mar96a} if one
takes into account  typical superdeformed radius parameters in the
penetrability calculations. The data were consistently interpreted
\cite{mar96a} in terms of production of a long-lived high spin
isomeric state in the second well of the potential energy surface
of $^{210}$Fr which decays, by a very enhanced
$\alpha$-transition, to a high-spin state in the second well of
$^{206}$At.

The predicted \cite{sat91,kri92} excitation energies of the second
minima in $^{210}$Fr and nearby nuclei are above the proton
separation energies \cite{aud93}. Therefore, the decay of isomeric
states from the second minima, by emitting protons, is in
principle possible. In a separate study of the same
$^{16}$O+$^{197}$Au reaction \cite{mar96b}, long-lived proton
radioactivities with half-lives of about 5.8 h and 67.3 h have
been discovered. They were interpreted as due to very retarded
decays from superdeformed isomeric states in the parent nuclei to
normal deformed or to the ground states of the daughters. In
particular, the indicated line with E$_{p}$ = 2.19 MeV
\cite{mar96b} may be associated with the predicted \cite{sat91}
(E$_{p}$ = 2.15 MeV) second minimum to ground state transition
from $^{198}$Tl, which can be produced by three consecutive
superdeformed to superdeformed $\alpha$-transitions from
$^{210}$Fr.

\section{Study of the $^{28}$S\lowercase{i}+$^{181}$T\lowercase{a}
reaction and long-lived
high spin hyperdeformed isomeric states}

The $^{28}$Si+$^{181}$Ta reaction has been studied at E$_{Lab}$ =
125 MeV, which is about 10\% below the Coulomb barrier, and at
E$_{Lab}$ = 135 MeV \cite{mar01a}. A fusion cross section of about
10 mb is predicted at 125 MeV using a coupled-channel deformation
code \cite{fer85} with deformation parameters $\beta_{2}$ = 0.41
for $^{28}$Si and $\beta_{2}$ = 0.26 for $^{181}$Ta \cite{ram87}
 and allowing for 2$^+$ and
3$^-$ excitations in $^{28}$Si. Only 2 $\mu$b is predicted when no
deformations are included in the calculations. For 135 MeV the
corresponding predicted fusion cross sections are 95 mb with
deformations and 40 mb without.

Figure 2 (left) shows an $\alpha$ - $\gamma$ two-dimensional
coincidence plot obtained at E$_{Lab}$ = 125 MeV. Quite a few
coincidence events are seen between a relatively high energy
$\alpha$-particle group around 8.6 MeV and various $\gamma$-rays.
The half-life of this coincidence group has been measured
\cite{mar01a} to be 40 d $\leq$ t$_{1/2}$ $\leq$ 2.1 y. Figure 2
(right) shows that the $\gamma$-rays which are in coincidence with
these high-energy $\alpha$-particles fit nicely with a J(J + 1)
law assuming E$_{x}$ = 4.42$\times$J(J +1) keV and $\Delta$J = 1.
Based on the observation of a Pt X-ray in coincidence with the 8.6
MeV $\alpha$'s and on kinematic arguments it was suggested
\cite{mar01a} that the $\alpha$-transition is from $^{195}$Hg to
$^{191}$Pt. ($^{195}$Hg may be produced via 1p1n evaporation
reaction and 3 consecutive III$^{min}$$\rightarrow$III$^{min}$
$\alpha$-decays. See below).
 An energy parameter of 4.42 keV is typical to
superdeformed band $\gamma$-ray transitions in this region of
nuclei.

\begin{widetext}

\begin{figure}[h]
\includegraphics[width=0.42\textwidth]{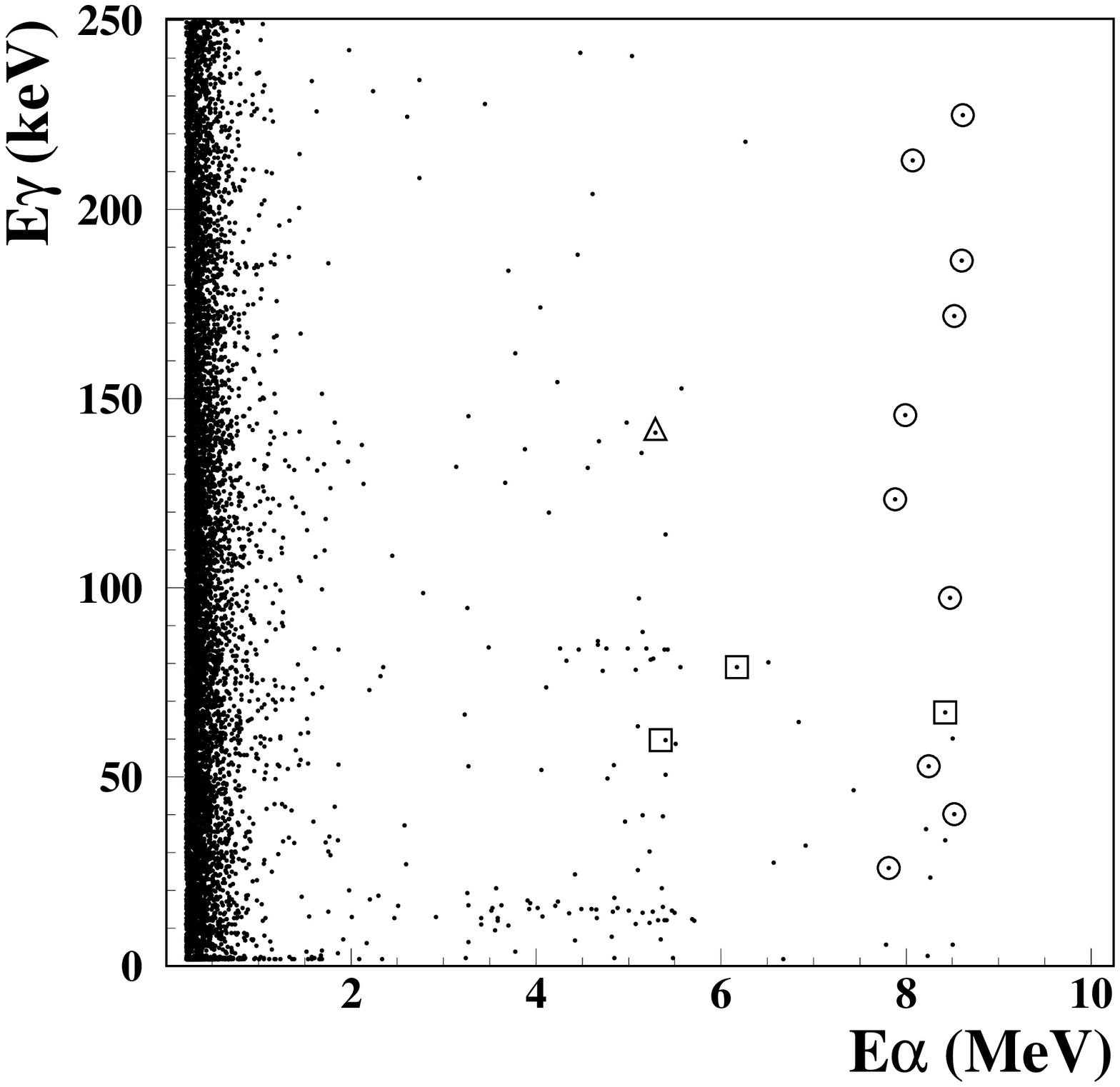}
\includegraphics[width=0.38\textwidth]{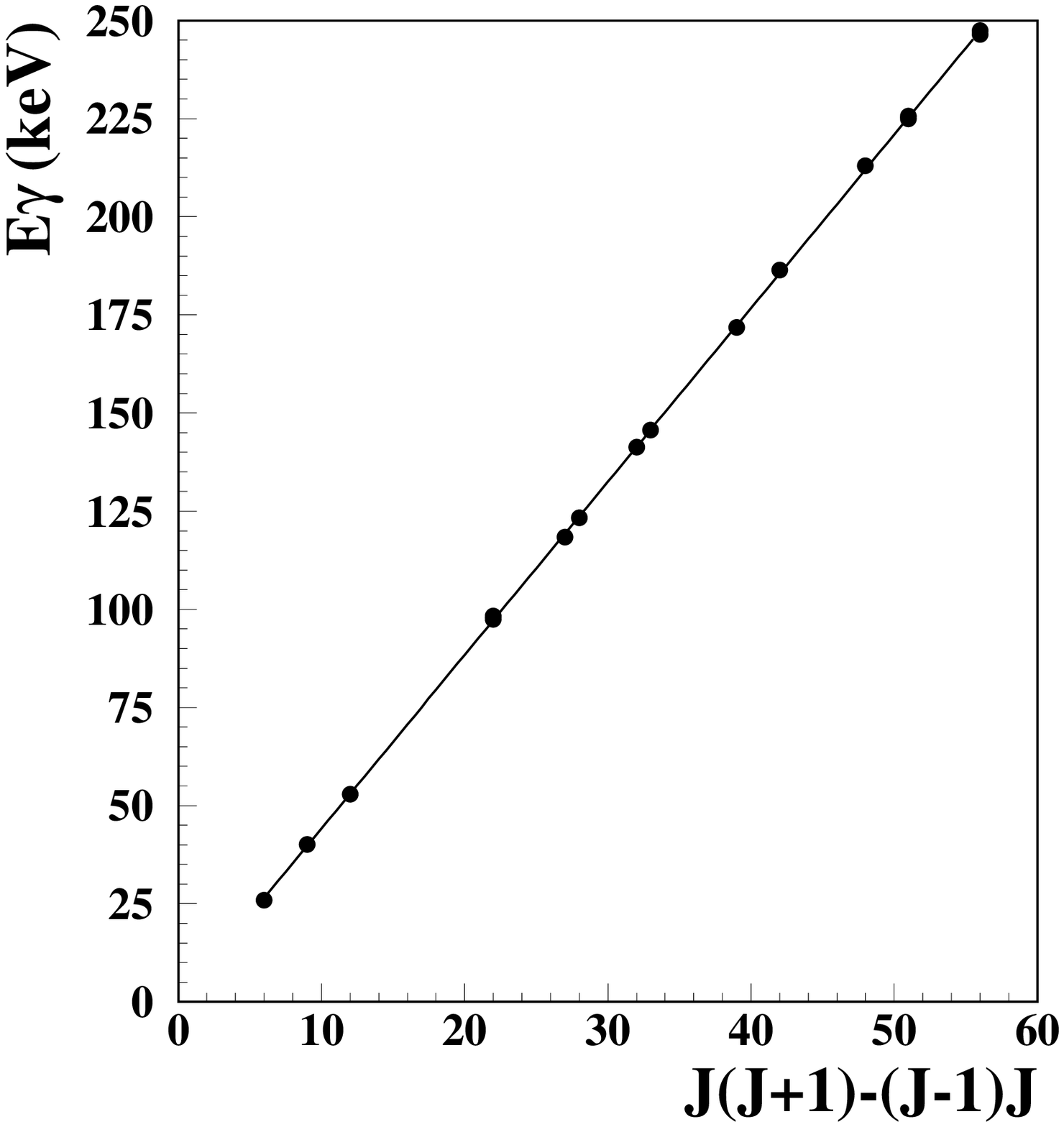}
\vspace*{-0.4cm}
 \caption{{\bf Left:}
$\alpha$ - $\gamma$ coincidence plot from one measurement of the
$^{28}$Si + $^{181}$Ta reaction. E$_{Lab}$ = 125 MeV, with 200
$\mu$g/cm$^{2}$ C catcher foil, taken for 76.8 d, starting 77.4 d
after the end of irradiation. The $\gamma$ - ray energies of the
encircled events fit with SDB transitions. The squared events fit
with known characteristic X - rays and the events in triangles are
identified with known $\gamma$ - ray transitions (see Ref.
\cite{mar01a}). {\bf Right:} E$\gamma$ versus J(J+1)-(J-1)J for
the $\gamma$ - rays seen in coincidence with 7.8 - 8.6 MeV
$\alpha$ - particles.
 (The encircled events in the left figure plus similar
events obtained in a second measurement \cite{mar01a}). The slope
of the straight line is 4.42 MeV.}
\end{figure}

\end{widetext}

An $\alpha$-energy of 8.6 MeV is a very high energy  for
$^{195}$Hg which does not decay by emitting $\alpha$-particles and
its g.s. to g.s. $Q_\alpha$-value is 2.190 MeV \cite{aud93}. A
half-life of 40 d is about 13 orders of magnitude too long as
compared to the systematics of energy versus half-life
relationship \cite{roy00} which predicts
t$_{1/2}\approx6\times10^{-8}$ s. Since the $\alpha$-particles are
in coincidence with a superdeformed band $\gamma$-ray transitions,
the $\alpha$-decay is to the superdeformed well of the daughter
nucleus. However, it could not be a
II$^{min}$$\rightarrow$II$^{min}$ transition, since such a
transition is very enhanced as opposed to the large
 retardation measured in the experiment. A consistent interpretation,
  both from the point of view of the high energy of the
  $\alpha$-particles and their very long lifetime, is that the
  decay is from a long-lived high spin (J$\approx$39/2) isomeric state in the
  III$^{min}$, the hyperdeformed minimum \cite{how80,naz93,cwi94,kra98}
   of $^{195}$Hg, which decays
   by strongly retarded transition to the II$^{min}$  of the potential
    in $^{191}$Pt \cite{mar01a}. As seen in Fig. 3 the predicted
    $Q_\alpha$-value for such a transition is about 8.7 MeV,
    taking into account an
    extrapolated value from Ref. \cite{naz93} for the excitation
    energy of the III$^{min}$ in $^{195}$Hg and the predictions of
    Refs. \cite{sat91,kri92} for the excitation energy of the
    II$^{min}$ in $^{191}$Pt. This value fits rather nicely with
    the measured $Q_\alpha$-value of about 8.8 MeV. (The
    excitation energy of the state in the third minimum of
    $^{195}$Hg was assumed to be around the rotational 39/2 state
    with estimated energy of E$_{x}$=2.2$\times$J(J +1) keV
    \cite{kra98}).

    \begin{figure}[h]
\includegraphics[width=0.42\textwidth]{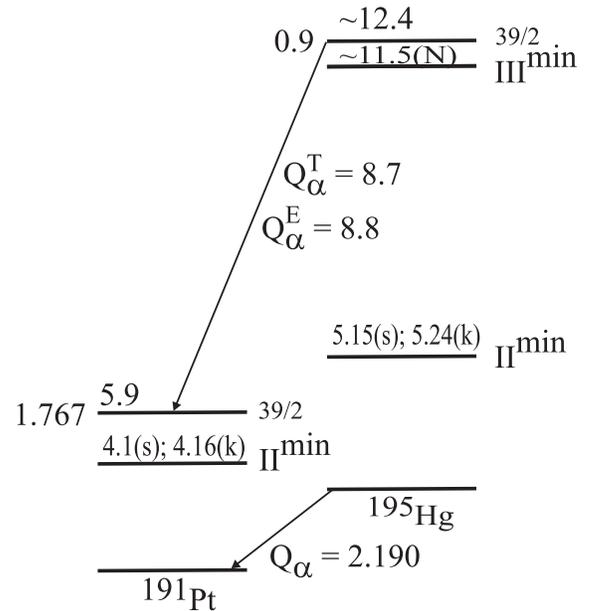}
\vspace*{-0.4cm}
 \caption{Proposed decay scheme deduced from the observation of the 8.6 MeV
 $\alpha$-particles seen in coincidence with a superdeformed band transitions.
 s - Satula et al. (Ref. \cite{sat91}); k - Krieger et al. (Ref.
 \cite{kri92}); N - Nazarewicz (Ref. \cite{naz93}).}
\end{figure}

\section{Super- and Hyper-deformed Isomeric States in the Actinide
Region}

Based on the discovery of the long-lived super- and hyper-deformed
isomeric states (Secs. III and IV), a consistent interpretation
has been given \cite{mar01b} to the unidentified $\alpha$-particle
groups seen in the actinide sources and described in Sec. II
above. II$^{min}$$\rightarrow$II$^{min}$ and
III$^{min}$$\rightarrow$III$^{min}$ $\alpha$-particle transition
energies have been deduced from the predicted \cite{how80}
excitation energies. Table II shows that the low energy 5.14, 5.27
and 5.53 MeV groups seen in the Bk, Es and No-Lr sources, can
consistently be interpreted as due to the low-energy
II$^{min}$$\rightarrow$II$^{min}$ transition from $^{238}$Am and
III$^{min}$$\rightarrow$III$^{min}$ transitions from $^{247}$Es
and $^{252}$No, respectively. Their energies are considerably
lower than the corresponding g.s. to g.s. transitions (Column 6 in
Table II).

\begin{table}[h]
\vspace*{-0.5cm} \caption[]{A comparision between the experimental
$\alpha$-particle energies and values deduced from the predictions
of Ref. \cite{how80} for some superdeformed (SD) to superdeformed
and hyperdeformed (HD) to hyperdeformed isomeric transitions. The
last column shows the corresponding experimental g.s. to g.s.
transitions \cite{aud93}. }
\begin{minipage}{0.5\textwidth} 
\renewcommand{\footnoterule}{\kern -3pt} 
\begin{tabular}{llllll}
\hline\\[-10pt]         
 Source &     E$_{\alpha}^{exp}$   &   Isotope   & Transition &
  E$_{\alpha}^{cal}$\cite{how80} &
 E$_{\alpha}^{g.s.{\rightarrow}g.s.}$\\
   &  (MeV)  &   &    & (MeV)    &  (MeV)\\
\hline\\[-10pt]
  Bk  &  5.14   & $^{238}$Am\footnote{Since the intensity of the 5.14 MeV group
(Fig. 1) grew at the beginning with time, it was assumed here
\cite{mar01b}, similar to the situation with the isomeric states
in $^{236}$Bk and $^{236}$Am \cite{mar87}, that $^{238}$Bk decayed
by EC or $\beta^{+}$ transitions to $^{238}$Am.}  &
SD$\rightarrow$SD & 5.13 &
  5.94\\
    Es  &  5.27   & $^{247}$Es  &   HD$\rightarrow$HD  & 5.27  &
  7.32\\
    No-Lr  &  5.53   &  $^{252}$No  &
  HD$\rightarrow$HD  &  $\approx$5.6\footnote{Extrapolated value.
     See Ref. \cite{mar01b}.}  & 8.42\\

\hline
\end{tabular}
\vspace*{0.2cm}

\end{minipage}
\renewcommand{\footnoterule}{\kern-3pt \hrule width .4\columnwidth
\kern 2.6pt}            

\vspace*{-0,3cm}

\end{table}



Table III shows that the very enhanced measured half-lives  of the
low-energy $\alpha$-particle groups seen in the various actinide
sources  are consistent with calculated values \cite{mar01b},
taking into account in the penetrability calculations the
deformation parameters of the superdeformed and hyperdeformed
isomeric states.

\begin{table}[h]
\vspace*{-0.5cm} \caption[]{Experimental and predicted half-lives
for superdeformed (SD) to superdeformed and hyperdeformed (HD) to
hyperdeformed transitions \cite{mar01b}.}

\begin{minipage}{0.5\textwidth} 
\renewcommand{\footnoterule}{\kern -3pt} 
\begin{tabular}{llllllll}
\hline\\[-10pt]         
Mother &     E$_{\alpha}$   &  Transition   &
$\beta_{2}$$\footnote{$\beta_{2}$ and $\beta_{4}$ values were
deduced from the $\epsilon_{2}$ and $\epsilon_{4}$ values given in
Ref. \cite{how80} using Fig. 2 of Ref. \cite{naz96}. The value of
$\beta_{3}$ was taken equal to $\epsilon_{3}$.}$ &
$\beta_{3}$$^{a}$ & $\beta_{4}$$^{a}$ &
t$_{1/2}^{cal}$\footnote{Calculated according to formulas given in
Ref. \cite{mar01b}. Calculated half-lives for other deformation
parameters are given in Ref. \cite{mar01b}.} &
t$_{1/2}^{exp}$\\
 Isotope  &  (MeV)  &    &   &    &    &  (y) &  (y) \\
\hline\\[-10pt]
 $^{238}$Am  &  5.14   & SD$\rightarrow$SD  & 0.71  &
   0.0  &  0.09  &  10.9 & 3.8$\pm$1.0 \\
 $^{247}$Es  &  5.27   & HD$\rightarrow$HD  & 1.05  &
   0.19  &  0.0  &  1.15 & 1.7$\pm$0.2 \\
 $^{252}$No  &  5.53   & HD$\rightarrow$HD  & 1.2  &
   0.19  &  0.0  &  0.22  & 0.07$\pm$0.02 \\
\hline
\end{tabular}
\end{minipage}
\renewcommand{\footnoterule}{\kern-3pt \hrule width .4\columnwidth
\kern 2.6pt}            
\end{table}

The  potential energies as function of quadrupole deformations,
taking from Ref. \cite{how80}, are shown in Fig. 4 for the
$^{238}$Am, $^{238}$Cm, $^{247}$Es, and $^{248}$Fm nuclei. It is
seen that in $^{238}$Am the inner and the outer barriers of the
second minimum are quite large, while in $^{247}$Es and $^{248}$Fm
the outer barriers of the second minima are  small, and the inner
barriers of the third minima are  large. In fact, the third minima
in $^{247}$Es and $^{248}$Fm are predicted to be the ground states
of these nuclei, being 0.61 and 1.76 MeV  below the normal,
slightly deformed, ground states. Unfortunately there are no
predictions in these cases for the potential at even larger
deformations, beyond the third minimum. (In the case of $^{232}$Th
\cite{cwi94} the outer barrier in the third minimum is quite
high).

\begin{figure}[h]
\vspace*{1.5cm}
\includegraphics[width=0.5\textwidth]{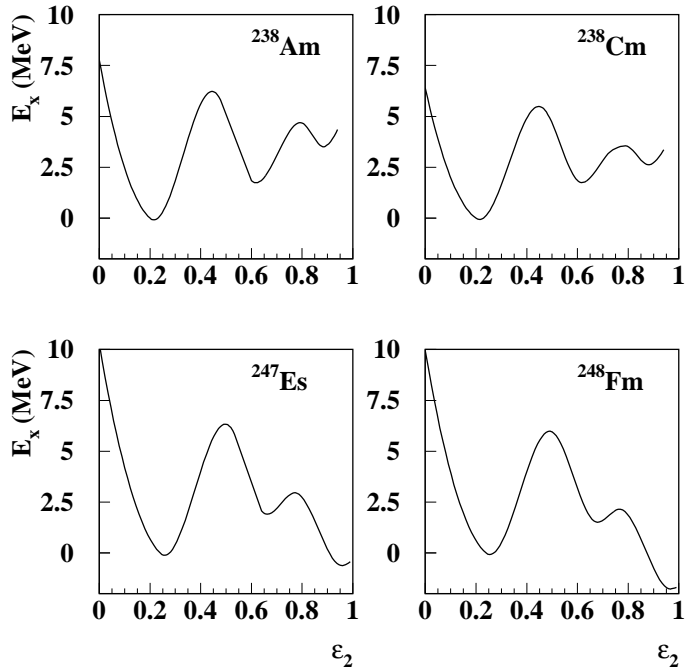}
\vspace*{-0.4cm}
 \caption{Potential energies as function of quadrupole deformations
for 4 nuclei according to Ref. \cite{how80}.}
\end{figure}

In Ref. \cite{mar01b} detailed estimates for the various
production cross sections of the actinide nuclei, as well as of
the superheavy element with Z = 112 and N $\simeq$ 160, are given.
It is argued that the relatively large fusion cross sections, in
the mb region, are due to two effects. First, the compound nucleus
is produced in an isomeric state in the second or third minimum of
the potential, rather than in the normal ground state. As shown in
Fig. 5 much less overlapping and penetration are needed under
these conditions, and therefore the compound nucleus formation
probability increases drastically. Secondly, in the secondary
reaction experiment the projectile is a fragment which has been
produced within 2$\times$10$^{-14}$ s before interacting with
another W nucleus in the target. During this short time it is at
high excitation energy and quite deformed. Figure 6 gives the
results of couple-channel calculations \cite{fer85} for the fusion
cross section as function of bombarding energy for the
$^{70}$Zn+$^{186}$W reaction, taking into account the known
deformation of $^{186}$W and  various deformations of the
projectile. Figure 6d shows the results when the projectile has a
deformation which is typical for the second minimum of the
potential. It is seen that, due to the reduced Coulomb repulsion
between the two nuclei for the tip to tip configuration, the cross
section reduces very slowly with decreasing bombarding energy.

An idea about the relative importance of the above two effects can
be deduced from the following arguments: The difference from a
typical cross section of about 1 pb \cite{hof96} obtained in the
$^{70}$Zn + $^{208}$Pb reaction producing the nucleus $^{277}$112
in its ground state, to a cross section of about 20 nb producing
$^{271(2)}$112 in an isomeric state via the $^{88}$Sr + $^{184}$W
reaction \cite{mar93}, is due to the first effect. The additional
difference from 20 nb to about 3.8 mb \cite{mar01b} of producing
element 112 in an isomeric state via secondary reactions, is due
to the second effect.

\begin{figure}
\hspace*{-0.6cm}
\includegraphics[width=0.54\textwidth]{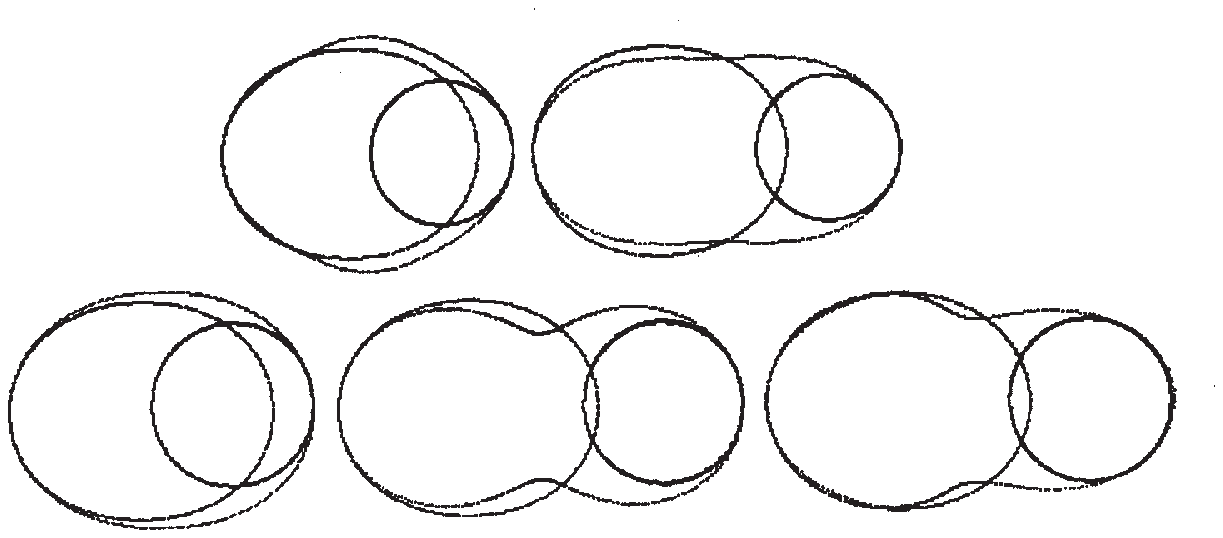}
\vspace*{-0.4cm} \caption{Calculated shapes of two compound nuclei
at various configurations together with the shapes of the
corresponding projectile and target nuclei. Top, left: A$_{C.N.}$
= 239 in the normal ground state; $\beta$$_{2}$ = 0.2;
$\beta$$_{4}$ = 0.08 \cite{how80}. Top, right: A$_{C.N.}$ = 239 in
the second minimum; $\beta$$_{2}$ = 0.77; $\beta$$_{4}$ = 0.1
\cite{how80}. In both figures: A$_{heavy}$ = 186; $\beta$$_{2}$ =
0.22 \cite{ram87}. A$_{light}$ = 53; $\beta$$_{2}$, $\beta$$_{3}$,
$\beta$$_{4}$ = 0.0. Bottom, left: A$_{C.N.}$ = 253 in the normal
ground state; $\beta$$_{2}$ = 0.28; $\beta$$_{4}$ = 0.01
\cite{how80}. Bottom, center: A$_{C.N.}$ = 253 in the third
minimum; $\beta$$_{2}$ = 1.2; $\beta$$_{4}$ = 0.0 \cite{how80}.
Bottom, right: A$_{C.N.}$ = 253 with parameters of the third
minimum of $^{232}$Th; $\beta$$_{2}$ = 0.85; $\beta$$_{3}$ = 0.35;
$\beta$$_{4}$ = 0.18 \cite{cwi94}. In the three figures at the
bottom: A$_{heavy}$ = 186; $\beta$$_{2}$ = 0.22 \cite{ram87}.
A$_{light}$ = 67; $\beta$$_{2}$, $\beta$$_{3}$, $\beta$$_{4}$ =
0.0.}
\end{figure}


\section{Summary of the properties of the super- and
hyper-deformed isomeric states}

Figure 7 summarizes the results obtained about the super- and
hyper-deformed isomeric states. The nucleus may have a long
lifetime in its ground state, but also in long-lived isomeric
states in the second and third minima of the potential-energy
surfaces. Long-lived superdeformed isomeric states may decay by
very enhanced $\alpha$-particles to superdeformed states in the
daughter nuclei, or by strongly retarded $\alpha$-particles to the
normal deformed states in the corresponding  nuclei. It also may
decay by very retarded proton radioactivity. Similarly,
hyperdeformed isomeric states may decay by strongly enhanced
$\alpha$-particle decay to the hyperdeformed potential well in the
daughter, or by very retarded $\alpha$-decay to the superdeformed
minimum in the same nucleus. All these extremely  unusual decay
properties have been discovered experimentally as summarized in
Fig. 7.

\begin{figure}
\includegraphics[width=0.45\textwidth]{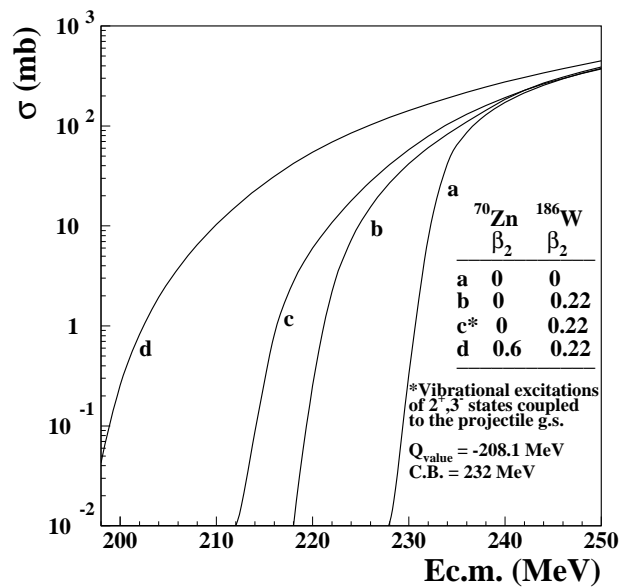}
 \caption{Calculated fusion cross sections using
the Code CCDEF \cite{fer85} for the $^{70}$Zn + $^{186}$W reaction
assuming various quadrupole deformations of the projectile and
target nuclei (see text, Sec. V).}
\end{figure}

It should be mentioned that the half-lives of the  newly
discovered isomeric states are longer than those of their
corresponding ground states. Such a comparison for the isomeric
states in the actinide region is presented in Table IV.

 It should be mentioned that back in 1969 \cite{nil69} a new type
of fission isomeric state has been predicted for nuclei with N
$\approx$ 144-150, A specialization energy in excess of 4 MeV for
the second barrier was predicted for a [505]$\frac{11}{2}^{-}$
state, which is associated with a factor of about 10$^{15}$
increase in the half-life of a normal fission shape isomer.

\begin{table}[h]
\caption {Half-lives of some isomeric states and their ratios to
the half-lives of their corresponding normal-deformed ground
states.}
\begin{minipage}{0.5\textwidth} 
\renewcommand{\footnoterule}{\kern -3pt} 
\begin{tabular}{lllc}
\hline
 Isotope & t$_{1/2}^{g.s.}$  & t$_{1/2}^{i.s.}$  &
t$_{1/2}^{i.s.}$/t$_{1/2}^{g.s.}$ \\ \hline
\\[-10pt]
$^{236}$Bk &  42.4 s\footnote{Predicted by P.~M\"{o}ller et al.
\cite{mnk}.} & $\geq$ 30 d\footnote{Ref. \cite{mar87}.} &  $\geq$ 6.1 x 10$^{4}$\\
$^{236}$Am &  3.6 m\footnote{Y. Nagame et al., this proceedings.}
 & 219 d$^{b}$  &   8.8 x 10$^{4}$\\
$^{238}$Am\footnote{Assuming that the 5.14 MeV is from $^{238}$Am
(see
Tables 1 and 3).} & 98 m\footnote{Ref. \cite{fir96}.}  & 3.8 y & 2.0 x 10$^{4}$\\
$^{247}$Es\footnote{Assuming that the 5.27 MeV is from
$^{247}$Es (see Tables 1 and 3).} & 4.55 m$^{e}$ & 625 d &  2.0 x 10$^{5}$\\
$^{252}$No\footnote{Assuming that the 5.53 MeV is from $^{252}$No
(see text).} & 2.3 s$^{e}$ & 26 d &  9.8 x 10$^{5}$\\
 \hline
\end{tabular}
\end{minipage}
\renewcommand{\footnoterule}{\kern-3pt \hrule width .4\columnwidth
\kern 2.6pt}            
\end{table}

\begin{widetext}

\begin{figure}
\mbox{
\begin{minipage}[h] {0.46\linewidth}
\includegraphics[height=6.6cm]{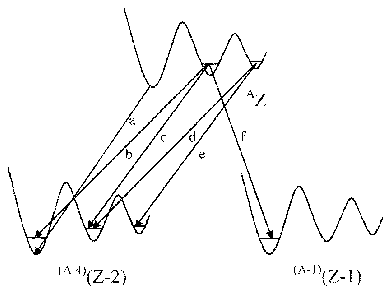}
\vspace*{0.5cm}
\end{minipage}\hfill

\begin{minipage}[h] {0.46\linewidth}
 a) I$^{min}$$\rightarrow$I$^{min}$. Normal
$\alpha$'s.\\
 b) II$^{min}$$\rightarrow$I$^{min}$. Retarded $\alpha$'s:\\
 \hspace*{2.0cm} $^{190}$Ir$\rightarrow$$^{186}$Re \cite{mar01a}.\\
  c) II$^{min}$$\rightarrow$II$^{min}$. Enhanced $\alpha$'s:\\
 \hspace*{2.0cm} $^{210}$Fr$\rightarrow$$^{206}$At \cite{mar96a}.\\
 \hspace*{1.0cm} $\sim$$^{238}$Am$\rightarrow$$^{234}$Np \cite{mar01b}.\\
  d) III$^{min}$$\rightarrow$II$^{min}$. Retarded $\alpha$'s:\\
 \hspace*{2.0cm} $^{195}$Hg$\rightarrow$$^{191}$Pt \cite{mar01a}.\\ e)
III$^{min}$$\rightarrow$III$^{min}$. Enhanced $\alpha$'s:\\
 \hspace*{2.0cm} $^{\sim247}$Es$\rightarrow$$^{243}$Bk \cite{mar01b}.\\
 \hspace*{2.0cm} $\sim$$^{252}$No$\rightarrow$$^{248}$Fm \cite{mar01b}.\\ f)
II$^{min}$$\rightarrow$I$^{min}$. Retarded protons
\cite{mar96b,mar01a}.\\
 \hspace*{2.0cm} $^{198}$Tl$\rightarrow$$^{197}$Hg(?) \cite{mar96b}.\\
 \hspace*{2.0cm} $^{205}$Fr$\rightarrow$$^{204}$Rn(?) \cite{mar01a}.\\
\end{minipage}
}
\vspace*{-0.2cm} \caption{Summary of abnormal particle decays seen
in various experiments}
\end{figure}

\end{widetext}

\section{Super- and hyper-deformed isomeric states and the
puzzling phenomena seen in nature}

The discovered super- and hyper-deformed long-lived isomeric
states enable one to understand the previously puzzling phenomena
seen in nature (See the Introduction).

The source for  the Po halos \cite{hen39,gen68} may be such
isomeric states in isotopes with Z $\simeq$ 84 which decayed, by
$\beta$- or $\gamma$-decays, to the ground states of $^{210}$Po,
$^{214}$Po and $^{218}$Po.

\begin{figure}[h]
\includegraphics[width=0.48\textwidth]{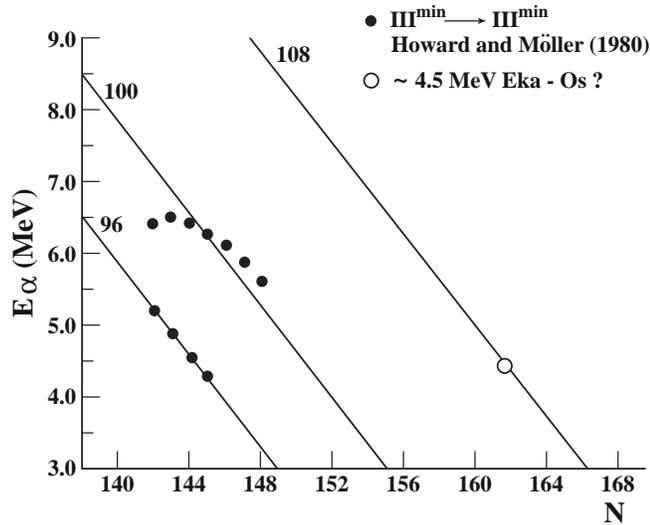}
\caption{Predictions \cite{how80}, and
extrapolations from these predictions, of the III$^{min}$
$\rightarrow$ III$^{min}$ $\alpha$-particle energies. The black
dots are the predictions for various isotopes of Z=96 and Z=100.
The straight lines are extrapolations from these predictions. The
open circle shows the position of 4.5 MeV $\alpha$-particles in
Z=108.}
\end{figure}


The low energy $\alpha$-particles around 4.5 MeV
\cite{cher63,che64,cher68,mei70} can consistently be interpreted
as due to a very enhanced III$^{min}$ $\rightarrow$ III$^{min}$
transition in Z$\sim$108 and A$\sim$271. The predicted
\cite{mar96a,mar01b} half-life in this case is around 10$^{9}$ y,
as seen in Table V. This resolves the first difficulty in
understanding these data, namely the about eight orders of
magnitude shorter lifetime than what is predicted
\cite{vs66,roy00} from energy versus lifetime relationship for a
normal $\alpha$-transition. (About $2.5\times10^{8}$ y estimated
experimentally  as compared to $5\times10^{16}$ y. See the
Introduction).

\begin{table}[h]
\caption{Calculated half-lives for hyperdeformed to hyperdeformed
$\alpha$-particle transition of 4.5 MeV from $^{271}$Hs assuming
various deformation parameters \cite{mar01b}.}
\begin{minipage}{0.5\textwidth} 
\renewcommand{\footnoterule}{\kern -3pt} 
\begin{tabular}{llll}
\hline
 $\beta_{2}$ & $\beta_{3}$ & $\beta_{4}$ &
t$_{1/2}$ (y)\\ \hline
\\[-10pt]
1.2\footnote{$\epsilon_{2}$ and $\epsilon_{4}$ values for
$^{248}$Fm were taken from Ref. \cite{how80} and converted to
$\beta_{2}$ and $\beta_{4}$ values according to Ref.
\cite{naz96}.} & 0.0\footnote{Assuming $\beta_{2}$ = 0.}
 & 0.0$^{a}$ & 1.8$\times$10$^{11}$\\
1.2$^{a}$ & 0.19\footnote{Assuming $\beta_{3}$ = $\epsilon_{3}$ of
Ref. \cite{how80}.}
  & 0.0 & 4.6$\times$10$^{9}$\\
0.85\footnote{Parameters given in Ref. \cite{cwi94} for
$^{232}$Th.}
 & 0.35$^{d}$  & 0.18$^{d}$ & 1.3$\times$10$^{8}$\\
\hline
\end{tabular}
\end{minipage}
\renewcommand{\footnoterule}{\kern-3pt \hrule width .4\columnwidth
\kern 2.6pt}            
\vspace{-0.3cm}
\end{table}

In addition, an extrapolation of the deduced $\alpha$-energies for
III$^{min}$$\rightarrow$III$^{min}$ transitions from the
predictions of Ref. \cite{how80} shows that for Z=108, E$_\alpha$
of about 4.5 MeV corresponds to N$\sim$162 (see Fig. 8). This is
consistent with the suggestion \cite{mei70} that $^{247}$Cm may be
a descendent of the superheavy element with Z=108 which decays by
the 4.5 MeV $\alpha$-particles, since $^{247}$Cm can be obtained
from $^{271}_{108}$Hs$_{163}$ by successive six $\alpha$-decays.
Another possibility is that the long-lived isotope is $^{267}$Hs
which decays by two $\beta^{+}$ or electron capture decays to
$^{267}$Sg which then follows by five successive $\alpha$-decays
to $^{247}$Cm. It should however be mentioned that in principle
the above 4.5 MeV $\alpha$-particles may  also be due to a
strongly retarded II$^{min}$$\rightarrow$I$^{min}$ or
III$^{min}$$\rightarrow$II$^{min}$ transition in the region of Os
itself. (For normal 4.5 MeV $\alpha$-particles in Os the expected
\cite{vs66,roy00} half-life is about 1 y. Such short-lived nuclide
can not exist in nature).

\section{Summary}

It was shown that the newly discovered long-lived super- and
hyper-deformed isomeric states can provide consistent
interpretations to two previously unexplained phenomena seen in
nature. Thus, the Po halos can be understood as due to the
existence of such isomeric states in nuclei with Z values around
84 and atomic masses in the region of 210 - 218. The observed 4.5
MeV $\alpha$-particle group can be understood as due to a low
energy and strongly enhanced hyperdeformed to hyperdeformed
transition in a nucleus with Z = 108 and A $\simeq$ 271.

It seems to us that the existence of superheavy elements in nature
is not impossible.

\section{Acknowledgements}

We appreciate very much the valuable discussions with J. L. Weil
and  N. Zeldes.


\begin{thebibliography}{99}
\bibitem{str64}V. M. Strutinskii, Yadernaya Fizika {\bf3,} 614 (1964).
\bibitem{mye66}W. D. Myers and W. J. Swiateski, Nucl. Phys. {\bf 81,} 1 (1966).
\bibitem{sob66}A. Sobiczewski, F. A. Gareev and B. N. Kalinkin, Phys. Lett.
 {\bf 22,} 590 (1966).
\bibitem{str67}V. M. Strutinskii, Nucl. Phys. {\bf A95,} 420 (1967).
\bibitem{won67}C. L. Wong, Phys. Rev. Lett. {\bf 19,} 328 (1967).
\bibitem{muz68}Yu. A. Muzychka, V. V. Pashkevich and Strutinskii,
Dubna Preprint R7-3733, 1968.
\bibitem{nil68}S. G. Nilsson, J. R. Nix, A. Sobiczewski, Z. Szymanski, S. Wycech, C.
Gustafson and P. M\"{o}ller, Nucl. Phys. {\bf A115,} 545 (1968).
\bibitem{gru69}J. Grumann, U. Mosel, B. Fink and W. Greiner, Z. Physik {\bf228,} 371
(1969).
\bibitem{hen39}G. H. Henderson and F. W. Sparks, Proc. Roy. Soc. Lond., {\bf
A173}, 238 (1939).
\bibitem{gen68}R. V. Gentry, Science {\bf160}, 1228 (1968).
\bibitem{gen92}R. V. Gentry, {\it Creation's Tiny Mystery}, Earth
Science Associates, Knoxville, Tennessee (1992).
\bibitem{cher63}V. V. Cherdyntsev and V. F. Mikhailov,
Geochemistry No. 1, 1 (1963).
\bibitem{che64}R. D. Chery, K. A. Richardson and J. A. S. Adams, Nature
{\bf 202}, 639 (1964).
\bibitem{cher68}V. V. Cherdyntsev, V. L. Zverev, V. M. Kuptsov and G. I.
Kislitsina, Geochemistry  No. 4, 355 (1968).
\bibitem{mei70}H. Meier et al., Z. Naturforsch. {\bf 25}, 79 (1970).
\bibitem{jol07}J. Joly, J. Phil. Mag. {\bf 13,} 381 (1907).
\bibitem{mug07}O. Mugge, Zent. Mineral. {\bf 1907,} 397 (1907).
\bibitem{fea78}N. Feather, Cumm. Roy. Soc. Edinburgh,  No. 11,
147 (1978).
\bibitem{mnk}P. M\"{o}ller, J. R. Nix and K. -L. Kratz, At.
Data Nucl. Data Tables {\bf66}, 131 (1997).
\bibitem{kuty}H. Koura, M.Uno, T. Tachibana and M.Yamada, Nucl.
Phys. {\bf A674}, 47 (2000); RIKEN-AF-NP-394 (April 2001).
\bibitem{lmz}S. Liran, A. Marinov and N. Zeldes, Phys. Rev.
{\bf C62}, 047301 (2000); arXiv:nucl-th/0102055.
\bibitem{vs66}V. E. Viola, Jr. and G. T. Seaborg, J. Inorg. Nucl.
Chem. {\bf28,} 741 (1966).
\bibitem{roy00}G. Royer, J. Phys. G: Nucl. Part. Phys. {\bf26,} 1149 (2000).
\bibitem{mar01b}A. Marinov, S. Gelberg, D. Kolb and J. L. Weil,
Int. J. Mod. Phys. {\bf E10,} 209 (2001).
\bibitem{mar71}A. Marinov, C. J. Batty, A. I. Kilvington, G. W. A.
Newton, V. J. Robinson and J. D. Hemingway, Nature {\bf 229,} 464
(1971).
\bibitem{mar96a}A. Marinov, S. Gelberg and D. Kolb,  Mod. Phys. Lett.
 {\bf A11,} 861 (1996).
\bibitem{mar96b}A. Marinov, S. Gelberg and D. Kolb,  Mod. Phys. Lett.
{\bf A11,} 949 (1996).
\bibitem{mar01a}A. Marinov, S. Gelberg and D. Kolb, Int. J. Mod. Phys.
{\bf E10,} 185 (2001).
\bibitem{mar02}A. Marinov, S. Gelberg, D. Kolb, R. Brandt and A.
Pape, To be published in the Proc. 3rd Int. Conf. on Exotic Nuclei
and Atomic Masses, July 2-7, 2001, H\"{a}meenlinna, Finland.
\bibitem{mar87}A Marinov, S. Eshhar and D. Kolb,  Phys. Lett.  {\bf B191,} 36 (1987).
\bibitem{sat91}W. Satula, S. \'{C}wiok, W. Nazarewicz, R. Wyss and
A. Johnson, Nucl. Phys. {\bf A529,} 289 (1991).
\bibitem{kri92}S. J. Krieger, P. Bonche, M. S. Weiss, J. Meyer,
 H. Flocard and P. -H. Heenen, Nucl. Phys. {\bf A542,} 43
(1992).
\bibitem{aud93}G. Audi and A. H. Wapstra, Nucl. Phys. {\bf A565,}
66 (1993).
\bibitem{fer85}J. Fern\'{a}ndez-Neillo, C. H.~Dasso and S.~Landowne,
   Code CCDEF, Comp. Phys. Comm.  {\bf 54,} 409  (1985).
\bibitem{ram87} S. Raman, C. H.~Malarkey, W. T.~Milner, C. W.~Nestor,~Jr.
 and P. H.~Stelson, At. Data and Nucl. Data Tables
 {\bf 36,} 1 (1987).
\bibitem{how80} W. M.~Howard and P.~M\"{o}ller, At. Data and Nucl. Data Tables
 {\bf 25,} 219  (1980).
\bibitem{naz93} W.~Nazarewicz, Phys. Lett.  {\bf B305,} 195  (1993).
\bibitem{cwi94} S.~\'{C}wiok, W.~Nazarewicz, J. X.~Saladin, W.~Pl\'{o}ciennik and
A.~Johnson, Phys. Lett.  {\bf B322,} 304 (1994).
\bibitem{kra98}A. Krasnahorkay et al., Phys. Rev. Lett. {\bf 80,}
2073 (1998).
\bibitem{naz96}W. Nazarewicz and I. Ragnarsson, Handbook of Nuclear
Properties, eds. D. N. Poenaru and W. Greiner (Clarendon Press,
Oxford, 1996) p. 80.
\bibitem{hof96}S. Hofmann et al., Z. Phys. {\bf A354,} 229 (1996).
\bibitem{mar93}A. Marinov, S. Gelberg and D. Kolb, Inst. Phys.
Conf. Ser. No. 132: Section 4, Nuclei Far from Stability/Atomic
Masses and Fundamental Constants 1992, p. 437.
\bibitem{fir96}R.~B.~Firestone, V. S. Shirley, C. M. Baglin, S. Y. F. Chu
and J. Zipkin,  Table of Isotopes (Wiley-Interscience, 1996).
\bibitem{nil69}S. G. Nilson, G. Ohl\'{e}n, C. Gustafson and P.
M\"{o}ller, Phys. Lett. {\bf 30B}, 437 (1969).

\end{thebibliography}
\end{document}